\documentclass[aps,prb,twocolumn]{revtex4}  
\usepackage{epsfig}
\usepackage{amsmath}
\usepackage{ulem}
\usepackage{color}
\newcommand{\DM}{Dzyaloshinskii-Moriya }
\begin{document}
\title{Electron spin resonance in $S=1/2$ antiferromagnets at high temperature}
\author{S. El Shawish,$^1$ O.~C\'epas,$^2$ and S.~Miyashita$^3$}
\affiliation{$1.$ Department of Theoretical Physics, Jo\v{z}ef Stefan Institute, SI-1000 Ljubljana, Slovenia.
 \\$2.$   Institut N\'eel, CNRS et Universit\'e Joseph Fourier, BP 166, F-38042 Grenoble 9, France 
\\ $3.$ Department of Physics, Graduate School of Science, The University of Tokyo, Bunkyo-Ku, Tokyo, 113033, Japan.
}

\begin{abstract}
We study the electron spin resonance (ESR) of low-dimensional spin
systems at high temperature, and test the Kubo-Tomita theory of
exchange narrowing. In finite-size systems (molecular magnets), we
found a double-peak resonance which strongly differs from the usual
Lorentzian.  For infinite systems, we have predictions for the
linewidth and lineshape as a function of the anisotropy strength. For
this, we have used an interpolation between a non-perturbative
calculation of the memory function at short times (exact
diagonalization) and the hydrodynamic spin-diffusion at long times. We
show that the Dzyaloshinskii-Moriya anisotropies generally induce a
much larger linewidth than the exchange anisotropies in two
dimensions, contrary to the one-dimensional case.
\end{abstract}

\pacs{PACS numbers:}
\maketitle
\date{\today}

\section{Introduction}

The paramagnetic resonance is a well-known phenomenon resulting from
the collective precession of the total magnetization about an external
magnetic field. For spin systems with interactions of the standard
SU(2) Heisenberg form, the total magnetization is conserved and there
is no relaxation of the magnetization, regardless of the strength of
the interaction. It is no longer true when anisotropic interactions
are present, and the magnetization relaxes with a characteristic time
scale. A theoretical issue is to relate this time scale (or the width
of the resonance) to the anisotropy strength in a many-body system.
In strongly interacting systems, small anisotropies (as in transition
metal compounds) lead to an exchange-narrowed Lorentzian
resonance.\cite{AndersonWeiss} The linewidth is given, generically, by
the theory of Kubo-Tomita,\cite{KuboTomita} and cast into a more
general formalism by Mori\cite{Mori} and Zwanzig.\cite{Zwanzig}
However, the Markovian assumption used by Kubo-Tomita was later argued
to break down in low dimensional systems.  For one dimensional
systems, deviations from the Lorentzian lineshape were indeed observed
experimentally and attributed to spin diffusion,\cite{Dietz,Benner}
which was not taken into account in the first place. Recently,
Oshikawa and Affleck\cite{OA} have developed a direct approach in
one-dimensional $S=1/2$ systems, based on an effective field-theory,
which is valid at low temperatures and could account successfully for
the (low-)temperature dependence of the measured linewidths. It is,
however, clear that a theory of the high-temperature linewidth would
be very useful to extract the anisotropies experimentally. The
Kubo-Tomita formula has been widely used in this regime but not
thoroughly tested.

It is nonetheless known that, even in the high-temperature regime, a
direct application of the Kubo-Tomita formula fails in the case of \DM
anisotropy in one dimension,\cite{Choukroun,OA1} which is first-order
in the spin-orbit coupling in $S=1/2$ systems. It predicts indeed a
linewidth varying like $D^2/J$, which was originally argued to be
responsible for the large linewidth observed in particular in
CuGeO$_3$.\cite{Yamada96,Eremina} It is in fact incompatible with a
general argument in one dimension, which leads to predict a smaller
$D^4/J^3$.\cite{Choukroun,OA1} It is then of the same order of
magnitude as the contribution of the exchange anisotropy (as also
happens for other observables\cite{KSEA}), \textit{i.e.} fourth-order
in the spin-orbit coupling and, hence, small. As alternatives,
unconventionnal superexchange\cite{Eremin} or dynamical \DM
interactions\cite{Dynamical} were invoked to explain the strong
broadening. Nonetheless, it is expected, on general grounds, that
one-dimensional systems may possess an inherent strong broadening
because of spin-diffusion at high temperatures. One issue is to obtain
a reliable quantitative estimation of the effect. Approximations, such
as the random phase approximation (RPA),\cite{Dietz} perturbative
expansions,\cite{Gulley} or self-consistent RPA,\cite{Reiter}
suggested in particular departures from the Lorentzian lineshape, but
it is difficult to judge how quantitative these theories are.

In contrast, in two spatial dimensions, it is surprising to notice
(empirically) that applying the Kubo-Tomita's formula seems to give
rather accurate results at high temperatures. For instance, in the
Shastry-Sutherland compound, SrCu$_2$(BO$_3$)$_2$, this gives an
out-of-plane \DM interaction $D=2.4$ K,\cite{Zorko0} while neutron
scattering gives 2.1 K.\cite{Kakurai} 

In the
following, we will give a quantitative estimation of the linewidth in
one and two dimensions and the corresponding lineshapes. We predict an enhancement of the linewidth with respect to the perturbative Kubo-Tomita formula, particularly strong when the diffusive motion is assumed at long times.  Moreover, while we agree
that the linewidth induced by the \DM anisotropy or the exchange
anisotropy are of the same order in one dimension (or more generally for
the \textit{reducible}\cite{Fai} \DM interactions), we claim that,
in two dimensions, the \textit{irreducible} \DM interactions lead to a
linewidth essentially varying like $D^2/J$, \textit{i.e.} like in Kubo-Tomita,
with a prefactor that we shall
estimate.

An approach consists of exact numerical diagonalizations of the
Hamiltonian and a calculation of the dynamics using the Kubo
formula.\cite{Miyashita} The approach is interesting not only because
there is no approximation made but also because it provides
information on the whole ESR spectrum and tells us whether there are
more resonances than the paramagnetic resonance. It is of course
limited by the system size and especially at finite temperatures where
one needs to calculate all eigenstates. As we shall see, a direct
computation of the Kubo formula gives results which are difficult to
extrapolate to the thermodynamic limit. Since, however, the
calculation is exact for small sizes, the prediction for the lineshape
may be interesting for molecular magnets, in particular to disentangle
different mechanisms. However, to obtain information for bulk systems,
we suggest that it is more interesting to calculate the memory
function by exact diagonalization, in the Mori-Zwanzig
framework. While the calculation of the memory function is exact only
at short-times for finite-size systems, its validity goes beyond the
perturbative short-time expansion developed
earlier.\cite{KuboTomita,Gulley} We find a clear slowing down of the memory function compared to Kubo-Tomita in one dimension (but not in two dimensions), possibly
indicative of a crossover to spin diffusion. We can then test the spin diffusion
assumption for the long-time behaviour.  It is indeed reasonable to
assume that the short-time behaviour depends on the microscopic
details of the lattice (and hence needs a precise calculation) while
the long-time may be more universal (and depends primarily on the
spatial dimension). This allows to obtain a quantitative idea of how
accurate the Kubo-Tomita's result is in one and two dimensions.

\section{Kubo formula and finite-size systems}
\label{section1}

The linear response of the spin system to a long wavelength oscillating magnetic field along, say, an $x$-axis is given by the Kubo formula (or susceptibility),
\begin{equation}
\chi(\omega)=  i  \int_{0}^{+\infty}  d\omega e^{i\omega t}  \langle [ S^x(t),S^x(0)] \rangle 
\end{equation}
where $S^x=\sum_i S_i^x$ is the total magnetization along the  $x$-axis.
 The absorption cross-section of the incident electro-magnetic wave is proportional to $\omega \chi''(\omega)$, and 
\begin{equation}
\chi''(\omega) = \pi (1-e^{-\beta \omega}) S^{xx}(\omega),
\end{equation}
where
\begin{eqnarray}
S^{xx}(\omega) &=& \int_{-\infty}^{+\infty} d\omega e^{i\omega t} \langle S^x(t)S^x(0) \rangle  \nonumber \\
 &=& \frac{1}{{\cal Z}} \sum_{n,p} e^{-\beta E_n} |\langle n|S^x|p\rangle|^2 \delta(\omega -E_p+E_n)
\label{matrixelement}
\end{eqnarray}
with a sum over the eigenstates $|n\rangle$ and
$|p\rangle$ of the Hamiltonian (${\cal Z}$ is the partition
function). At high temperatures ($T \gg J$),  $\omega \chi''(\omega)=(\omega^2/T)S^{xx}(\omega)$ and we show only $S^{xx}(\omega)$ in the
following. Furthermore, all the states have
the same weight ($e^{-\beta E_n} \sim 1$) and $S^{xx}(\omega)$
is an even function of $\omega$ ($S^{xx}(\omega)=S^{xx}(-\omega)$).  To compute Eq.~(\ref{matrixelement}), we have calculated
all eigenenergies and eigenstates  of the
Hamiltonian by exact numerical diagonalization. We have first considered the Heisenberg Hamiltonian for one
dimensional spin rings (with periodic boundary conditions) with two
types of SU(2)-symmetry-breaking anisotropies, one is the \DM
interaction (first-order in spin-orbit coupling),
\begin{equation}
{\cal H} = \sum_i  J \textbf{\mbox{S}}_i . \textbf{\mbox{S}}_{i+1}  + \textbf{\mbox{D}}_i . ( \textbf{\mbox{S}}_i \times \textbf{\mbox{S}}_{i+1})   ,
\label{ham0}
\end{equation}
where $\mathbf{D}_i$ is a staggered vector from bond to bond (strenght $D$) and taken along the $z$-axis. The second is the $XXZ$ exchange anisotropy (second-order in spin-orbit coupling),
\begin{equation}
{\cal H'} = \sum_i  J \textbf{\mbox{S}}_i . \textbf{\mbox{S}}_{i+1}  -\delta J  S_i^z S_{i+1}^z ,
\label{ham1}
\end{equation}
with an easy-plane anisotropy ($\delta >0$). 
Importantly, ${\cal H}$ can be mapped onto ${\cal H'}$ by rotating
the spin operators, $\tilde{S}_i^{+}=e^{i(-1)^i\theta}S_i^+$ with $\theta=D/2J$, giving $\delta \equiv D^2/2J^2$ at the lowest order.\cite{KSEA,OA1,Choukroun} In this mapping, the long wavelength oscillating magnetic field acquires a staggered component. In the following, we restrict
ourselves to an external field $h$ along $z$. In this
case, the $z$-component of the total magnetization, $S^z$, is conserved and the
magnetic field splits the excited states according to $-hS^z$ but does not change the matrix elements in Eq.~(\ref{matrixelement}).  In addition, the operator $S^x$ changes $S^z$ by $\pm 1$, so that $E_{p}-E_n$
is shifted by $\pm h$. As soon as $h$ is larger than the linewidth,
the $h \neq 0$ lineshape is the same as for $h=0$ shifted by $\pm h$.

\begin{figure}[tbp]
\centerline{
 \psfig{file=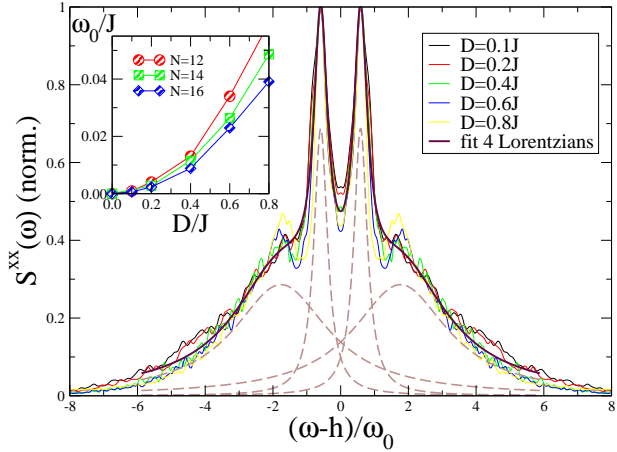,width=7.cm,angle=-90}}
\caption{(Color online) ESR lineshape of a one-dimensional chain of $N=16$ spins  ($T \gg J$). The curves are rescaled with the half-width at half-maximum, $\omega_0$ (inset). }
\label{Sxx}
\end{figure}

The result of the calculation of the Kubo formula
(\ref{matrixelement}) for $N=16$ spins interacting with the
Hamiltonian (\ref{ham0}) is shown in Fig.~\ref{Sxx} (the delta peaks
in (\ref{matrixelement}) are slightly broadened by a
Gaussian form\cite{artificialbroadening}). We find a broad resonance
centered at the magnetic field $h$ (there is no finite shift). Since we are at
high temperatures, we stress that $S^{xx}(\omega)$ is the same around $\omega=-h$. The calculation being exact, this is the lineshape we predict for
molecular magnets of the ring geometry, at high temperature, would the
mechanism giving the linewidth be purely magnetic in origin.
In Fig.~\ref{Sxx}, the different curves correspond to different
strengths of \DM coupling. They are normalized in intensity and
rescaled in frequency using the linewidth, noted $\omega_0$, and defined as the half-width at half-maximum of the full line.
They all collapse onto a single
(quasi)-universal lineshape with a double-peak structure, in a wide
range of anisotropy strengths, $D \sim 0.1-0.8 J$ and sizes. If we consider the special case $h=0$, the double-peak structure can be seen as a pseudo-gap.  The prediction of this lineshape seems at odds with the conventional
exchange-narrowing theory which, in general, predicts a single
Lorentzian. We will discuss the origin of these differences below.  In
addition to the double-peak structure, there is an additional weight in
the wing. In fact, the line-shape can be fitted by four Lorentzians,
two centered at $\pm 0.59\omega_0$ and two at $\pm 3(0.59\omega_0)$
(each component is given by a dashed line in Fig.~\ref{Sxx} and the
fit by a solid thick line). 

 The linewidth $\omega_0$ increases basically like $\alpha D^2/J$ (inset
 of Fig.~\ref{Sxx}) with a rather small prefactor, $\alpha$. The prefactor is size-dependent and is difficult to extrapolate to the
 thermodynamic limit. For this reason (and following the discussion of
 section \ref{KTMZ}), the special double-peak line-shape may or may not survive in the thermodynamic limit. For molecular rings, this should be relevant and the prefactor is found to be $\alpha=0.090$ ($N=12$), $0.071$ ($N=14$) and $0.059$ ($N=16$).

\begin{figure}[t]
\centerline{
 \psfig{file=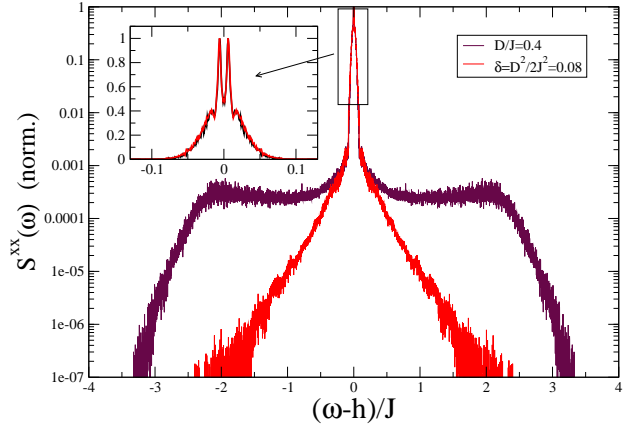,width=7.cm,angle=-90}}
\caption{(Color online) ESR lineshape (high-energy tail) for a one-dimensional chain ($T \gg J$) for the \DM and XXZ models (using $2\delta=D^2/J^2$). The response is identical only at low-energy (inset).}
\label{Background}
\end{figure}

We now emphasize the difference between the \DM model and the XXZ
model, as far as the response is concerned. In Fig.~\ref{Background}, we show
the resonance line in absolute values for the two models with coupling strengths related by the mapping $2\delta=(D/J)^2$. The two resonance lines are
undistinguishable at low energies (inset of Fig.~\ref{Background}) but
differ at high energy in the background (Fig.~\ref{Background}). The difference comes from 
the small oscillating staggered field (or umklapp contribution from the
spin fluctuations at $q=\pi$) that the mapping induces,\cite{Choukroun,OA1}
\begin{equation}
S^{xx}(\omega) = \cos^2\theta \tilde{S}^{xx}(\omega) +  \sin^2 \theta \tilde{S}^{xx}_{u}(\omega) 
\end{equation}
where $S^{xx}$ and $\tilde{S}^{xx}$ correspond to the \DM 
and XXZ model, respectively, and $\tilde{S}^{xx}_{u}(\omega)$ is the $q=\pi$ response. The second term gives extra ``forbidden''
resonances (or ``satellite'' lines) at low temperatures.\cite{Sakai}
 These resonances have disappeared at high temperatures. Instead,
this contribution gives a broad background, covering the whole energy
range. This is the origin of the difference of the two curves at high energy. To summarize, the resonance lineshape as shown in Fig.~\ref{Sxx} is the same for both  the \DM and the XXZ models as long as we use the correspondence $2\delta=(D/J)^2$ (and not too small intensities).

Note that we have found deviations
 about the universality of the low-energy lineshape at strong anisotropies (either \DM or XXZ). This is
 expected because, for $\delta=1$, the model is XY and the lineshape
 is known to be exactly Gaussian.\cite{Brandt} We have recovered this
 result numerically: we have found no difference with a Gaussian at
 $\delta=1$ (no finite-size effects). The way it crossovers from a
 Gaussian function to four Lorentzians (by reducing $\delta$ from 1) is
 to depress the response at $\omega=h$ and let side peaks appear,
 which decrease in height. It is therefore clear that in the range of
 strong anisotropies, the lineshape is not universal at all.

\section{Kubo-Tomita theory, exact calculation of the memory function}
\label{KTMZ}

We now discuss the Kubo-Tomita theory, in order to understand why the lineshape is not a single Lorentzian in such finite-size systems and go beyond the limitations imposed by finite-size calculations. A way to clarify the assumptions is to start with the more general Mori-Zwanzig formalism.\cite{Mori,Zwanzig} In this framework, the spin fluctuations (those that do not depend on time through the magnetization function) are viewed as a random noise (the bath) that forces the magnetization to equilibrate. The system is then governed by a Langevin equation,
\begin{equation}
\frac{d}{dt} S^{x} (t) = -  \int_{0}^t d\tau \Sigma(t-\tau) S^{x}(\tau) +\eta(t) \equiv F(t),
\label{diffequation}
\end{equation}
where the ``memory function''  (or self-energy) $\Sigma(t)$ is related to
 the correlation function of the random force $\eta(t)$ by the fluctuation-dissipation theorem. However, neither $\Sigma(t)$ nor $\eta(t)$ are easily expressed in terms of the original spin Hamiltonian because of unknown projection operators.\cite{Mori,Zwanzig} It is only at second-order in perturbation theory (here in the anisotropy),\cite{Zwanzig} that $\Sigma(t)$ is given by the total force correlation function
\begin{equation}
\Sigma(t) = \langle F(t) F^{\dagger}(0) \rangle 
\label{correlationfunction}
\end{equation}
where the total force (or torque) is given by
 $F(t)=i e^{i{\cal H}t} [{\cal H},S^x]e^{-i{\cal H}t}$. Since $F(t)$ is linear in the anisotropy, $\Sigma(t)$ is second-order, and perturbation theory holds for small anisotropies. Eq.~(\ref{correlationfunction}) is precisely what 
 Kubo and Tomita have obtained directly by perturbation theory.

The second assumption is that $S^x(t)$ is slow compared to the
relaxation time of the force (local equilibrium).\cite{Zwanzig} It is true that the
force $F$ is not conserved by the dynamics (whereas $S^x(t)$ is nearly
conserved) and evolves on a short time-scale of order 
$1/J$,\cite{AndersonWeiss,KuboTomita} but the absence of memory at long times is an assumption.  If it is true (the system is Markovian),  Eq.~(\ref{diffequation}) 
simply defines a relaxation time, $\tau_c$
(or Onsager's transport coefficient)
\begin{equation}
\frac{1}{\tau_c} = \int_{0}^{+\infty}  d\tau \Sigma(\tau).
\label{result}
\end{equation}
 In this ``fast'' regime, the lineshape is therefore always Lorentzian with
half-width at half-maximum (linewidth) given by $1/\tau_c$. In contrast, if slow hydrodynamic modes are present, the spin motion may be diffusive.\cite{Dietz,Benner} In this case, the memory function has a power-law tail $\Sigma(t) \sim 1/t^{d/2}$ and Eq.~(\ref{result}) diverges for $d \leq 2$. In fact, this signals a change of lineshape and the general solution of (\ref{diffequation}) for the correlation function is
\begin{equation}
S^{xx}(\omega) = \frac{1}{4\pi} \frac{\Sigma''(\omega)}{[\omega+\Sigma'(\omega)]^2+\Sigma''(\omega)^2}
\label{generalsolution}
\end{equation}
with $\Sigma(\omega)= i \int_{0}^{+\infty} \Sigma(t) e^{i\omega t
}dt$ the Laplace transform of the memory function, and $\Sigma'(\omega)$, $\Sigma''(\omega)$ are its real and imaginary parts. If ``slow''
processes are present, $\Sigma(\omega)$ varies considerably
near $\omega=0$ and the lineshape deviates from a Lorentzian. The form
of $\Sigma(t)$ is therefore of  importance and various
assumptions have been used. The original (third) assumption of Kubo-Tomita was to
use a Gaussian decay for $\Sigma(t)$, based on the short-time
evolution given by perturbation theory
\begin{equation}
\Sigma(t)= \sum_n \frac{(it)^n}{n!} M_{n+2} 
\label{st}
\end{equation}
where $M_{n+2}$ are the moments, $M_2=\Sigma(0)=\langle F^2 \rangle$,
$M_3= \langle [{\cal H},F] F \rangle$, $M_4=\langle [{\cal H},F] [F,{\cal H}] \rangle$
etc.. They can be calculated exactly at high temperatures.
For the XXZ model (\ref{ham1}), $M_2/J^2=\delta^2 /4$, $M_3=0$, and
$M_4/J^4=3\delta^2/8-\delta^3/4+\delta^4/4$.  So at short times, $\Sigma(t)/\Sigma(0)=1-\frac{1}{2!} \omega_e^2 t^2 $ with the definition $ \omega_e^2 \equiv \frac{M_4}{M_2} $.
Kubo and Tomita have postulated that
\begin{equation}
\Sigma(t) =\Sigma(0) \exp(-\omega_e^2 t^2/2)
\label{Sigmat}
\end{equation}
with $\omega_e^2$ matching that of the short-time expansion (\ref{st}) at second order, $\omega_e^2\equiv M_4/M_2=3J^2/2- \delta J^2 + \delta^2 J^2$. Eq.~(\ref{Sigmat}) then basically corresponds to assuming a ressummation of an infinite number of terms. The Kubo-Tomita formula for the linewidth is then at high temperature,
\begin{equation}
  \frac{1}{\tau_c} = \sqrt{\frac{\pi}{2}} \frac{M_2}{\omega_e} \approx  \sqrt{\frac{\pi}{3}} \frac{\delta^2}{4}J
\label{KT}
\end{equation}
To summarize, the assumptions that lead to Eq.~(\ref{KT}) are (i) perturbation theory in $\delta
 \ll 1$, (ii) Markovian behaviour, and
 (iii) perturbation theory in time $t\ll 1/J$.

\begin{figure}[tbp]
\centerline{
 \psfig{file=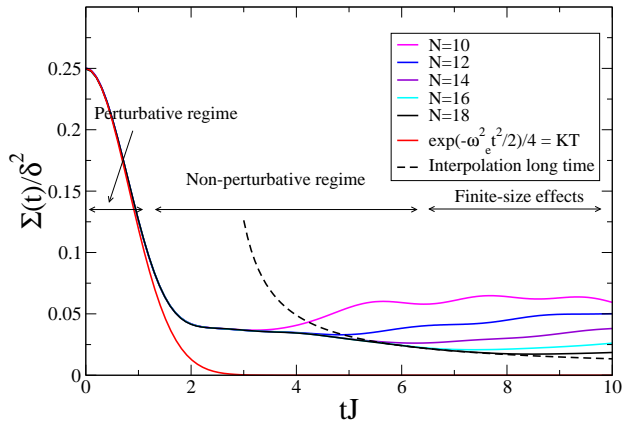,width=7.cm,angle=-90}}
\caption{(Color online) Memory function $\Sigma(t)$ at short times ($N$ is the system size,  $\delta=0.08$). Convergence to $N=\infty$ is obtained for $tJ \lesssim 6.5$. The Kubo-Tomita gaussian holds for $tJ \lesssim1.5$. The interpolation to the long-time spin-diffusion $1/t^{1/2}$ is shown (dashed line).}
\label{sigma}
\end{figure}

\begin{figure}[htbp]
\centerline{
 \psfig{file=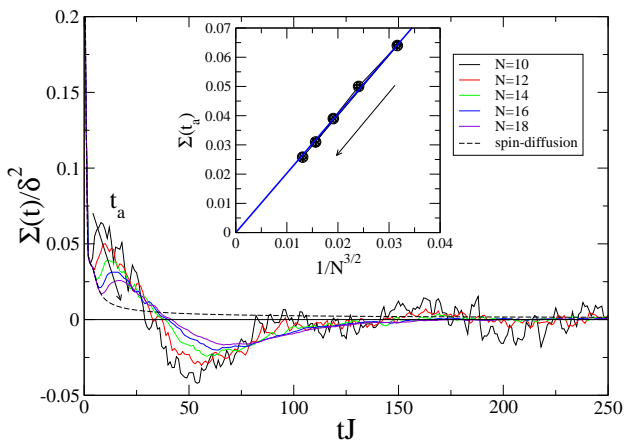,width=7.cm,angle=-90}}
\caption{(Color online) Memory function $\Sigma(t)$ at long times (system size $N$, $\delta=0.08$). There is a peak at $t_aJ\sim 10-17$ whose amplitude decreases with $N$ (inset). The assumed long-time spin-diffusion tail is also shown (dashed line).}
\label{sigmalongtime}
\end{figure}

In order to go beyond perturbation theory and test these assumptions, we have calculated
\begin{equation}
\Sigma(t) = \frac{1}{{\cal Z}} \sum_{n,p} e^{-\beta E_n} |\langle n |[{\cal H},S^x]| p \rangle|^2 e^{i(E_n-E_p)t}
\end{equation}
by exact diagonalization of the Hamiltonian [see Eq.~(\ref{matrixelement}) for the definitions].
We restrict ourselves again to high temperatures ($e^{-\beta E_n} =1$) and have computed all matrix elements. Here we calculate quantities for the model (\ref{ham1}) which makes no difference as we mentioned.
 At short times
(Fig.~\ref{sigma}), $\Sigma(t)$ decreases indeed on a time scale $\sim
1/J$ and the Kubo-Tomita gaussian decay holds for
$tJ \lesssim 1.5 $ (perturbative regime). For $tJ>1.5$, $\Sigma(t)$
decreases slowly and by comparing the results for different sizes, we see no difference between the $N=16$ and $N=18$ curves for $tJ \lesssim 6.5$. We can therefore be
confident that for $tJ \lesssim 6.5$, $\Sigma(t)$ is representative of the
thermodynamic limit. The result is different from the
gaussian decay and there is a considerable slowing down for $1.5 \lesssim t \lesssim 6.5$. We have here access to a regime that would be difficult to access by perturbation theory. We will come back on this non-perturbative regime in the next section. Above $tJ \gtrsim 6.5$, finite-size effects dominate. In particular at longer times
(Fig.~\ref{sigmalongtime}), $\Sigma(t)$ acquires an additional broad
peak at about $t_aJ \sim 10-17$. $t_a$ increases almost linearly with
the length of the chain $N$, and its amplitude decreases with $ \sim 1/N^{3/2}$ (as shown by the fit in the inset of
Fig.~\ref{sigmalongtime}). A possible interpretation is that this additional peak is related to the
torque travelling around the ring. In this case, indeed the arrival
time is expected to be controlled by $J$ and weakly sensitive to the
anisotropy, which is what we found. In fact if we take the des Cloizeaux-Pearson speed of the hydrodynamic excitation, the arrival time is $t_a=N/c=2N/\pi \sim 6-11$.
This additional peak together with the long-time decay  should be at the origin of the ``double-peak''
structure observed in Fig.~\ref{Sxx}. A simple way to see the effect of a long-time decay of $\Sigma(t)$ is to consider
the gaussian memory function of Eq.~(\ref{Sigmat}) with a long-time scale $1/\omega_e \gg 1/J$. The Laplace transform, $\Sigma(\omega)$, is a sum of a Gaussian function (imaginary part) and a Dawson function (real part), and $S^{xx}(\omega)$ [Eq.~(\ref{generalsolution})] acquires a double peak
structure.\cite{Abragam} This occurs when the time scale of the memory function becomes comparable to that of the spin motion $1/\omega_e \geq \tau_c\sqrt{2\pi}/(2+\sqrt{\pi})$. For a general $\Sigma(t)$, the change of sign of the curvature of Eq.~(\ref{generalsolution}) at $\omega=0$ signals the occurence of a double-peak structure. This happens when
$\tau_c \leq \langle t\rangle + \langle t^2 \rangle^{1/2}$
where $\langle t^n \rangle=(1/n!)\int_0^{+\infty} \Sigma(t)t^n dt/\int_0^{+\infty} \Sigma(t) dt$. As a consequence, there may be a critical size above which the double-peak structure disappears. These long-memory  processes (non-Markovian) explain why deviations from a Lorentzian lineshape take place in finite-size systems.

\section{Resonance in one dimensional thermodynamic systems}
\label{1d}

In the previous sections, we have given the exact lineshape of a
finite-size system. As we have explained, it is difficult to
extrapolate it in the thermodynamic limit. We now discuss a way to
obtain the resonance line in the thermodynamic limit by using
some additional assumptions. 
The crucial point is to
obtain the memory function
$\Sigma(t)$, but it cannot be computed at any time $t$ by exact
diagonalization. It is therefore interesting to assume a long-time behavior and
test its consequences on the resonance line, which can then be
compared to experiments. For this, we will interpolate between the exact
result at short times (up to the time where $\Sigma(t)$ has converged for $N \rightarrow \infty$, i.e. $tJ=6.5$ in Fig.~\ref{sigma}) and the assumed hydrodynamic behavior at long times.  

\begin{figure}[t]
\centerline{
 \psfig{file=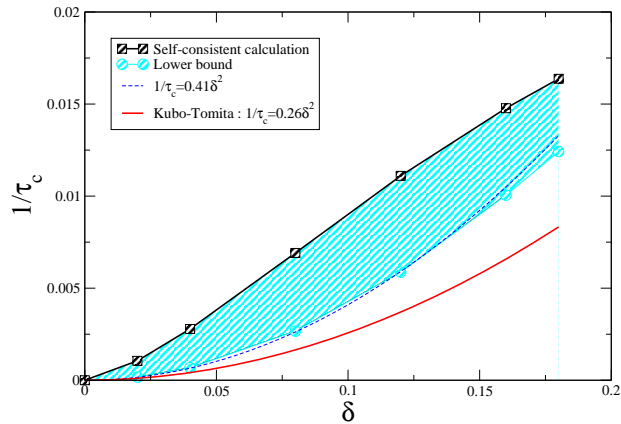,width=7.cm,angle=-90}}
\caption{(Color online) Linewidth vs. anisotropy for infinite one-dimensional chains (lower bound, Kubo-Tomita and self-consistent calculation).}
\label{linewidth}
\end{figure}

First we note that the slowing down obtained in the non-perturbative
regime ($1.5 \lesssim tJ \lesssim 6.5$) already leads to increase the
linewidth compared with Kubo-Tomita. If we assume for instance an
abrupt cutoff at $tJ=6.5$ (remember that for $tJ>6.5$, $\Sigma(t)$ is
dominated by finite-size effects), Eq.~(\ref{result}) defines a lower
bound for the linewidth. The lower bound, fitted by $\sim
0.41\delta^2$, is yet larger than the Kubo-Tomita formula,
$0.26\delta^2$ [Eq.~(\ref{KT})] (both shown in Fig.~\ref{linewidth}).

 We now consider the idea that the long-time behavior may be universal and governed by a spin diffusion equation.  We recall that the diffusion of $\Sigma(t)$
  is obtained assuming a RPA decoupling of the 4-spin correlations in
 $\Sigma(t)$,\cite{Dietz,Benner} $\langle S^+_q(t) S^z_q(t) S^z_{q'}(0)
 S^-_{q'}(0) \rangle \sim \langle S^+_q(t) S^-_{q'}(0) \rangle \langle
 S^z_q(t) S^z_{q'}(0) \rangle $. Since $S^z$ is conserved, the
 long wavelength modes are supposed to be diffusive with $\langle
 S^z_q(t) S^z_{-q}(0) \rangle \sim  e^{-D
   q^2t}$. On the other hand, $\langle S^+_q(t) S^-_{-q}(0) \rangle  \sim  e^{-D q^2t}e^{-t/\tau_c}$ is cut-off by the
 anisotropy, where $\tau_c$ is precisely the characteristic time-scale we are looking for, as emphasized by Reiter and Boucher.\cite{Reiter} When summed over $q$, this leads at long times ($t \gg1/J$)
 to
\begin{equation} \Sigma(t) \sim \frac{e^{-t/\tau_0}}{t^{d/2}}
\label{longtime}
\end{equation}
where $\tau_0$ is the cut-off time of the memory function: within the self-consistent RPA, $\tau_0 = \tau_c $. As explained above, we cannot test this behavior quantitatively by exact
diagonalization, although the slowing down for $1.5 \lesssim tJ \lesssim 6.5$ seems to indicate a crossover regime. To test the idea of spin-diffusion, Fabricius and McCoy have examined directly the two-spin
autocorrelation function for the $N=16$ chain at high temperatures and
concluded that the exponent may be closer to 0.7 in one dimension and
increases with $\delta$.\cite{Fabricius} So the exponent $1/2$ may be
underestimated, thus leading to overestimate the linewidth. Since $\tau_c$ now explicitely enters in Eq.~(\ref{longtime}), Eq.~(\ref{result}) is a self-consistent equation. This is precisely the equation that Reiter and Boucher solved.\cite{Reiter} Instead of relying on RPA to
obtain the prefactors, here we shall assume a long time tail of the form
(\ref{longtime}) and interpolate to the short-time behavior we have
obtained by exact diagonalization. The idea of interpolating the two types of
behavior was used by Gulley \textit{et al.}, together with
perturbative short-time expansion.\cite{Gulley} The interpolation
provides a ``variational'' function $\Sigma(t,\tau_c)$ and we solve (in a
similar spirit as in Ref.~\onlinecite{Reiter}) the (now)
self-consistent Eq.~(\ref{result}) with the new $\Sigma(t,\tau_c)$. In fact the long-tailed $\Sigma(t)$
changes the lineshape according to Eq.~(\ref{generalsolution}). In this case, Eq.~(\ref{result}) does not
hold anymore and we calculate the linewidth using instead the self-consistent equation, 
$S^{xx}(1/\tau_c)=S^{xx}(0)/2$ where $S^{xx}(\omega)$ is the solution given by Eq.~(\ref{generalsolution}). The difference between the two procedures
is in fact quite small (from 4\% to 7\% depending on $\delta$), and we show only the
full self-consistent result in Fig.~\ref{linewidth} (squares). This provides an estimation of the linewidth for a pure one dimensional model within the spin diffusion assumption.  

\begin{figure}[t]
\centerline{
 \psfig{file=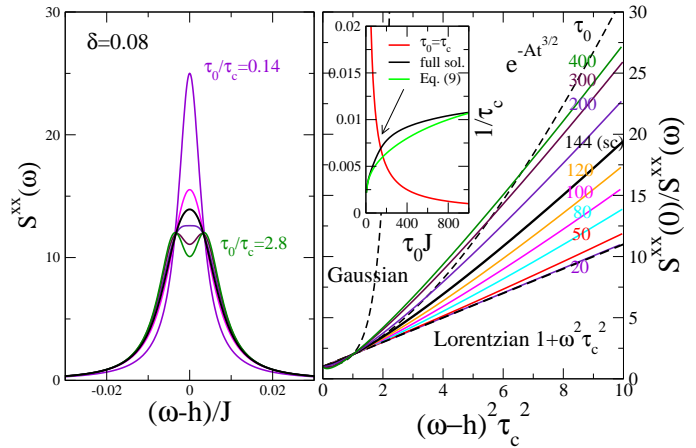,width=7.cm,angle=-90}}
\caption{(Color online) ESR lineshape for infinite one-dimensional chains, assuming different cutoffs  $\tau_{0}$ of the diffusive tail. The arrow (inset) indicates the self-consistent $\tau_c$.}
\label{lineshape}
\end{figure}

It is now possible to relax the assumption that the cutoff $\tau_0=\tau_c$ in
Eq.~(\ref{longtime}). 
It does not seem to be physically reasonable to take $\tau_{0}>\tau_c$ at high temperatures because it would mean having finite
torque-torque correlations but no spin-spin correlation after a
certain time.\cite{Notenematic} In
this case, Eq.~(\ref{result}) leads to a divergence of the
linewidth $\sim \sqrt{\tau_0}$ whereas the use of the correct lineshape provides a less diverging result (shown in the inset of
Fig.~\ref{lineshape}). The other assumption $\tau_{0}<\tau_c$ is
physically more relevant for two reasons. First $\tau_0$ may be an intrinsic characteristic time of the one-dimensional model that terminates the spin-diffusion before the anisotropy cut-off: $\tau_0=\tau_c$ is valid only at the RPA level. Second, $\tau_0$ may be an extrinsic ``noise'': there are additional interactions, such as
interchain couplings, that also tend to terminate the diffusion.\cite{Richardsinterchain,Boucher,Soos} In fact it
was argued that $\tau_{0}J$ depends on the interchain coupling $J'$
through $(J'/J)^{-4/3}$, which may indeed be a shorter time-scale.\cite{Richardsinterchain} On the other hand, following the exact result at short times, the
cutoff cannot be chosen smaller than $\tau_0J=6.5$ (but of course if $J'/J$
becomes very strong, then one has to recalculate the short-time
behavior as well).  The linewidth decreases when $\tau_0$ decreases; as a consequence, we regard the
region represented in Fig.~\ref{linewidth} by a hatched area as
possibly relevant, experimentally. As we now explain, in this region, the
lineshape is a function of $\tau_{0}$, thus giving the possibility to
say how long the motion is diffusive and if it is diffusive at all.

In Fig.~\ref{lineshape}, we give the line-shapes (normalized by the
area), computed from Eq.~(\ref{generalsolution}), for different
cut-offs of spin-diffusion, $\tau_{0}$. At small $\tau_0 \ll \tau_c$,
the lineshape is always Lorentzian (straight dashed line in the
inverse representation on the right). For the self-consistent cut-off
$\tau_0=\tau_c$, the lineshape strongly deviates from a Lorentzian. The deviation is less than that found by using the RPA calculation with a
spin-diffusion tail [which is given by the Fourier transform of
  $\exp(-At^{3/2})$ (see intermediate dashed line)]\cite{Dietz} and
still far from the gaussian profile (upper dashed line). For $\tau_0 > \tau_c$ we find a ``double peak'' structure similar to that found in finite-size chains. We have therefore two possibilities for the double-peak structure as a function of the system-size: either it survives in the thermodynamic limit (which means that for some reasons $\tau_0$ could remain larger than $\tau_c$) or it disappears. For a general
$\tau_0$, the inverse lineshape changes smoothly as function of
$\tau_0$ and therefore a comparison of the lineshape with experiment would provide a
measure of the cutoff time, $\tau_0$.

The linewidth (Fig.~\ref{linewidth}), together with the
lineshape (Fig.~\ref{lineshape}), can therefore be used to extract the
anisotropy strength and test the spin diffusion assumption. Note that
in one dimension, because of the mapping we have discussed, it is not
possible to tell whether the linewidth is due to the \DM or exchange
anisotropies:\cite{Choukroun,OA1} both contribute to an equal amount,
using the replacement $2\delta \equiv (D/J)^2$.

\section{Resonance in two dimensional thermodynamic systems}
\label{2d}

In two dimensions, the pattern of \DM vectors may generally be more
complicated, depending on the symmetries of the crystal structure. In
particular it is not always possible to perform the mapping onto an
exchange anisotropy because of the frustration that the closed loops
of the lattice introduce (\textit{irreducible components}). In this case, there seems to be no reason
why the \DM term should contribute to the linewidth by an equal amount
as the exchange anisotropy, and we shall see that it does not.

\begin{figure}[tbp]
\centerline{
 \psfig{file=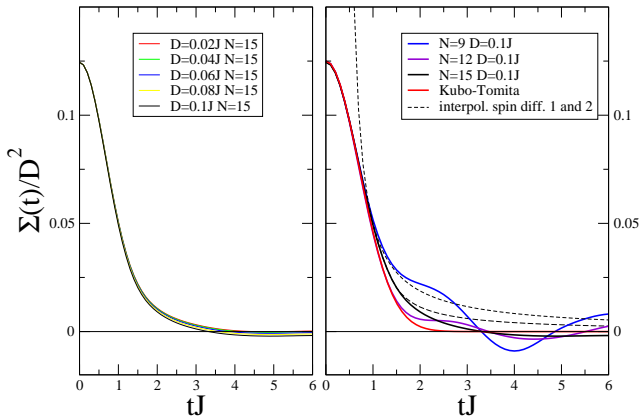,width=7.cm,angle=-90}}
\caption{(Color online) Memory function for the two-dimensional $S=1/2$ kagome antiferromagnet: effect of the \DM strength (left) and finite-size effects (right).}
\label{memory2d}
\end{figure}

We have considered the example of the kagome antiferromagnet where the pattern of \DM vectors precisely have irreducible components.\cite{CepasKagome} These are the $z$-component of the \DM field. We have calculated the memory function $\Sigma(t)$ by exact diagonalizations of clusters of up to 15 spins. We see that the result is weakly dependent on the strength of the \DM coupling, $D$ (Fig.~\ref{memory2d}, left), except through $\Sigma(0)=D^2/8$ (at high temperatures). As a consequence, the linewidth should  vary essentially like $D^2/J$. We now proceed like in one dimension to obtain a quantitative estimation of the prefactor. In fact finite-size effects are somehow more difficult to handle because we have less sizes available. In addition, we see a parity effect between clusters of odd and even sizes (Fig.~\ref{memory2d}, right). If we had only the $N=12$ cluster, it would be tempting to conclude that Kubo-Tomita is essentially exact in two dimensions. However, the $N=15$ cluster deviates from Kubo-Tomita. It is not clear whether this is an artifact due to the odd size of the cluster. We now proceed  with the interpolation to the spin diffusion assumption (which is in $1/t$ in two dimensions). Since we see no clear slowing down in the present case, we have used two different times for the interpolation, $tJ= 1$ (which certainly gives an upper bound) or $tJ=1.5$  (see the two dashed lines in Fig.~\ref{memory2d}).

\begin{figure}[htbp]
\centerline{
 \psfig{file=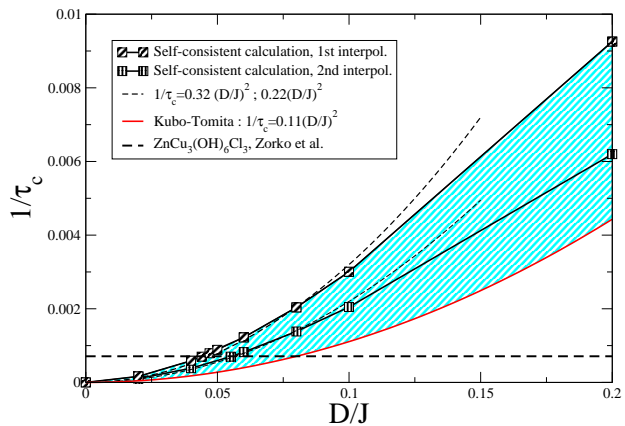,width=7.cm,angle=-90}}
\caption{(Color online) Linewidth vs. anisotropy for the  $S=1/2$ kagome antiferromagnet (Kubo-Tomita and self-consistent calculation). The horizontal line shows the experimental linewidth of ZnCu$_3$(OH)$_6$Cl$_3$ (Ref.~\onlinecite{Zorko}).}
\label{hwhm2d}
\end{figure}
\begin{figure}[htbp]
\centerline{
 \psfig{file=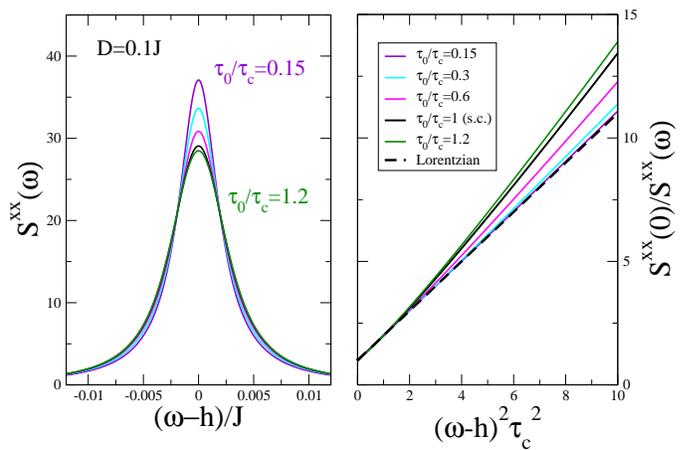,width=7.cm,angle=-90}}
\caption{(Color online) ESR lineshape for the two-dimensional kagome antiferromagnet, assuming different cutoffs $\tau_0$ of the $1/t$ diffusive tail}.
\label{lineshapes2d}
\end{figure}

The two results of the self-consistent calculation are shown in
Fig.~\ref{hwhm2d} (squares) and are fitted by $1/\tau_c=0.32 (D/J)^2$ and  $0.22(D/J)^2$
for small anisotropy.  The linewidth is therefore larger than in one
dimension. To put numbers, if we take $D=0.2J$, the linewidth is
$6.0 \times 10^{-3}J$ (Fig.~\ref{hwhm2d}), whereas for the
same $D$ in one dimension, we have to take $\delta=(D/J)^2/2=0.02$, and
the linewidth is $1.0 \times 10^{-3}J$ (Fig.~\ref{linewidth}). Now, as in one
dimension, the area below the self-consistent result in
Fig.~\ref{hwhm2d} could be experimentally relevant (because of
interplane couplings, for instance). Away from the Kubo-Tomita line
(lower bound in Fig.~\ref{hwhm2d}), we can see in
Fig.~\ref{lineshapes2d} that the lineshape starts to deviate from a
Lorentzian (but less than in one dimension).

\begin{figure}[htbp]
\centerline{
 \psfig{file=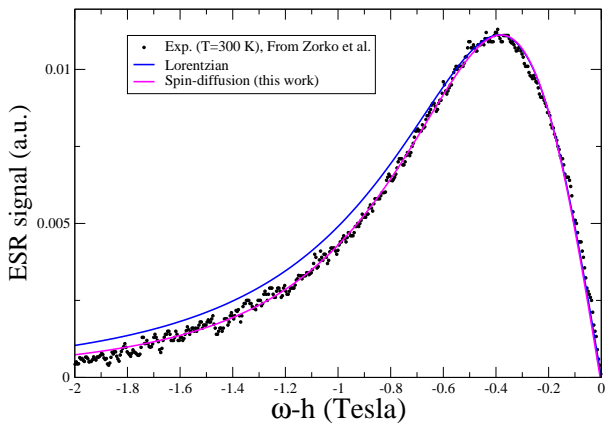,width=7.cm,angle=-90}}
\caption{(Color online) ESR signal (derivative of the resonance line at $\omega<h$): theory and experiment on the kagome $S=1/2$ ZnCu$_3$(OH$_3$)Cl$_2$ (from Ref.~\onlinecite{Zorko}).}
\label{explineshapes}
\end{figure}

We now apply this to the $S=1/2$ kagome compound
ZnCu$_3$(OH)$_6$Cl$_3$.\cite{Mendels}  The analysis of the linewidth using the
Kubo-Tomita formula provided $D/J=0.08$.\cite{Zorko} If we assume that
spin-diffusion takes place, the self-consistent calculation reproduces
the same experimental linewidth providing $0.044 \leq D/J \leq 0.08$
 (see dashed line
in Fig.~\ref{hwhm2d}).  It is difficult to compare directly with the ESR lineshape of ZnCu$_3$(OH)$_6$Cl$_3$ because the experiment was done on powder samples.\cite{Zorko} One needs to average over all directions of the field and we have assumed a single direction here. Nonetheless, Fig.~\ref{explineshapes} shows that the deviations from a
Lorentzian in the wings (here we restrict to $\omega<h$, which is not
spoiled by an impurity line) can be accounted for by the spin diffusion assumption and $D/J \sim 0.05$. The result may be partly coincidental because there are other sources of deviations, such as the chemical disorder and the polycristalline nature of the sample.
Still we note that this coupling strength is
not incompatible with the analysis of NMR which stated that
$D/J\gtrsim 0.05$.\cite{Rousochatzakis} In any case, from the ESR linewidth alone, we can conclude that $0.044 \leq D/J \leq 0.08$.

\section{Conclusion}

First, for finite-size chains (0 dimension), we have found a special (and exact) ``double-peak'' lineshape (Fig.~\ref{Sxx}) which may be interesting in molecular magnets having the geometry of a ring, especially because the lineshape seems insensitive to the strength anisotropy (up to a rescaling) and size (in the limit of small size). We have stressed that such a lineshape usually results from non-Markovian correlations in the memory function, which we did find in the calculation of $\Sigma(t)$ in this case. 

In the thermodynamic limit, we have obtained the electron spin resonance in  one and two
dimensions for the anisotropies relevant to $S=1/2$ systems, by computing the memory function. We have provided the
exact numerical result for $\Sigma(t)$ at short times and used it as
an initial condition for a long-time spin-diffusion tail, which
was taken as an assumption. We have then calculated the characteristic
relaxation time (or linewidth) self-consistently, as well as the lineshape.

In one dimension, the spin-diffusion assumption leads to enhance the
linewidth by a factor that we have quantitatively
estimated (see Fig.~\ref{linewidth} [squares]). It is about 4 times
larger than the perturbative Kubo-Tomita's result.  When this is true,
the lineshape strongly differs from a Lorentzian. To compare with
experiments, it is interesting to consider another time-scale, $\tau_0$ which is the cutoff of the memory function (the physical reason may be intrinsic or extrinsic; for instance, because of interchain
couplings). When $\tau_0$ is reduced from its
self-consistent value, the linewidth decreases and the lineshape
crossovers to a Lorentzian. A comparison of both the
linewidth and  lineshape should therefore allow experimentally to determine the
anisotropy strength and $\tau_0$ (with a better accuracy than the use of the
Kubo-Tomita formula).  Note that, given a resonance line, we can extract the
anisotropy parameter $\delta$ but we cannot tell whether $\delta$ is due to the
exchange anisotropy or the \DM field (unless the later is forbidden by
symmetry): both contribute at the same order of
magnitude in one dimension.\cite{Choukroun,OA1}

In two dimensions, the situation is different and there are two distinct cases:
\begin{itemize}
\item  \textit{Irreducible case} (the \DM vectors do
not sum to zero while going around closed loops of the lattice): we have argued that the linewidth should vary essentially like $D^2/J$.
\item \textit{Reducible case} (the \DM
vectors do sum to zero): an exact transformation  maps again the model onto an exchange
anisotropy and, as in one dimension, both contribute at the same order
 ($\delta^2J\sim D^4/J^3$).
\end{itemize}
The result in the irreducible case is in fact what the Kubo-Tomita formula  gives, qualitatively. Quantitatively, the prefactor is bounded below by
the Kubo-Tomita result. In fact if we restrict to the cluster of even
size ($N=12$), we see an excellent agreement with the gaussian decay
and we would be tempted to conclude that we see no
\textit{quantitative} difference with Kubo-Tomita. By computing the
next cluster ($N=15$), however, the difference turns out to be
larger. While it is difficult to know if the difference comes from the odd size of the cluster, we have given in the later case the
interpolation to the $1/t^{d/2}$ ($d=2$). The self-consistently
calculated linewidth is in this case close to $0.22D^2/J$ or $0.32D^2/J$
(Fig.~\ref{hwhm2d}) depending on the interpolation time, but still larger 
than the Kubo-Tomita's formula,
$0.11D^2/J$ (and in between, if interplane couplings are present).

We have applied the present theory to the two-dimensional kagome
ZnCu$_3$(OH$_3$)Cl$_2$,\cite{Mendels} and concluded that $0.044 \leq
D/J \leq 0.08$ from the ESR linewidth. In addition, the spin diffusion
assumption seems to account for the deviations of the lineshape from a
Lorentzian (Fig.~\ref{explineshapes}), but single crystals are needed to avoid additional averaging effects. In this compound, the
out-of-plane \DM component is irreducible while the in-plane is
reducible.\cite{CepasKagome} For single crystals, we therefore predict a much larger
linewidth when the field is perpendicular to the plane.

We also note that, in absolute values, the irreducible \DM interactions in two dimensions
induce a linewidth larger than in one-dimension (where the \DM
interactions are always reducible). 
The essential factor that governs the linewidth in
low-dimensional $S=1/2$ systems is not the strength of the divergence
$1/t^{1/2}$ versus $1/t$ of the spin-diffusion (assuming it exists) but rather the
reducible versus irreducible character of the \DM interaction. We have
illustrated this on the kagome lattice but, at high temperatures, the
shape of the lattice should not matter (as long as there are closed
loops with \DM vectors not summing to zero).

\section*{Acknowledgments}
We would like to thank A. Zorko for providing us with the experimental data of the lineshape (Ref.~\onlinecite{Zorko}).


\begin{thebibliography}{99}
\bibitem{AndersonWeiss} P.~W. Anderson and P.~R. Weiss, Rev. Mod. Phys. \textbf{25}, 269 (1953). 
\bibitem{KuboTomita} R.~Kubo and K.~Tomita, J. Phys. Soc. Jpn. \textbf{9}, 888 (1954).
\bibitem{Mori} H.~Mori, Progr. Theor. Phys. \textbf{33}, 423 (1965).
\bibitem{Zwanzig} R. Zwanzig, Phys. Rev. \textbf{124}, 983 (1961).
\bibitem{Dietz} R. E. Dietz, F. R. Merritt, R. Dingle, Daniel Hone, B. G. Silbernagel, and Peter M. Richards, Phys. Rev. Lett. \textbf{26}, 1186 (1971). 
\bibitem{Benner} H. Benner and J.-P. Boucher, in \textit{Magnetic Properties of Layered Transition Metal Compounds}, ed. L. J. de Jongh (Kluwer, Dordrecht, 1990) p. 323.
\bibitem{OA} M.~Oshikawa and I.~Affleck, Phys. Rev. Lett. \textbf{82}, 5136 (1999).
\bibitem{Choukroun} J.~Choukroun, J.-L.~Richard, and A.~Stepanov, Phys. Rev. Lett. \textbf{87}, 127207 (2001).
\bibitem{OA1} M.~Oshikawa and I.~Affleck, Phys. Rev. B  \textbf{65}, 134410 (2002).
\bibitem{Yamada96} I. Yamada, M. Nishi, and J. Akimitsu, J.Phys. Condens. Matt. \textbf{9}, 2625 (1996).
\bibitem{Eremina}
R. M. Eremina, M. V. Eremin, V. N. Glazkov, H.-A. Krug von Nidda, and A. Loidl, Phys. Rev. B \textbf{68}, 014417 (2003).
\bibitem{KSEA} T. Kaplan, Z. Phys. B: Condens. Matter \textbf{49}, 313 (1983); L. Shekhtman, O. Entin-Wohlman, and A. Aharony, Phys. Rev. Lett. \textbf{69}, 836 (1992).
\bibitem{Eremin} M. V. Eremin, D. V. Zakharov, R. M. Eremina, J. Deisenhofer, H.-A. Krug von Nidda, G. Obermeier, S. Horn, and A. Loidl, Phys. Rev. Lett. \textbf{96}, 027209 (2006).
\bibitem{Dynamical} M. V. Eremin, D. V. Zakharov, H.-A. Krug von Nidda, R. M. Eremina, A. Shuvaev, A. Pimenov, P. Ghigna, J. Deisenhofer, and A. Loidl, Phys. Rev. Lett. \textbf{101}, 147601 (2008).
\bibitem{Gulley} J. E. Gulley, Daniel Hone, D. J. Scalapino, and B. G. Silbernagel, Phys. Rev. B \textbf{1}, 1020 (1970). 
\bibitem{Reiter} G.~F.~Reiter and J.-P.~Boucher, Phys. Rev. B \textbf{11}, 1823 (1975).  
\bibitem{Zorko0} A.~Zorko, D. Ar$\breve{\mbox{c}}$on, H. van Tol, L. C. Brunel, and H. Kageyama, Phys. Rev. B \textbf{69}, 174420 (2004).
\bibitem{Kakurai} O. C\'epas, K. Kakurai, L. P. Regnault, J.-P. Boucher, T. Ziman, N. Aso, M. Nishi, H. Kageyama, and Y. Ueda, Phys. Rev. Lett. \textbf{87}, 167205 (2001).
\bibitem{Fai} Y. F. Cheng, O. C\'epas, P. W. Leung and T. Ziman, Phys. Rev. B \textbf{75}, 144422 (2007).
\bibitem{Miyashita} S. Miyashita, T.~Yoshino, and A.~Ogasahara, J. Phys. Soc. Jpn. \textbf{68}, 655 (1999);  A.~Ogasahara, and S. Miyashita, J. Phys. Soc. Jpn. \textbf{72}, 44 (2003).
\bibitem{artificialbroadening} For a finite-size system, one could argue that there is no need for broadening as it is the exact result once the Hamiltonian is given. In fact there are many additional sources of natural broadening. Even if these were absent, experimentalists usually modulates the external field by a (very) small-frequency component, thus artificially broadening each peak.
\bibitem{Brandt} U. Brandt and K. Jacoby, Z. Phys. B \textbf{25}, 181 (1976).
\bibitem{Sakai} T. Sakai, O. C\'epas, and T. Ziman, J. Phys. Soc. Jpn. \textbf{69}, 3521 (2000).
\bibitem{Abragam} A. Abragam and M. Goldmann, \textit{Nuclear Magnetism: Order and Disorder.} (Clarendon Press, Oxford), page 30.
\bibitem{Fabricius} K. Fabricius and B.~M.~McCoy, Phys. Rev. B \textbf{57}, 8340 (1998).
\bibitem{Richardsinterchain} Michael J. Hennessy, Carl D. McElwee, and Peter M. Richards, Phys. Rev. B \textbf{7}, 930 (1973).
\bibitem{Boucher} J.-P.~Boucher, M.~Ahmed Bakheit, M.~Nechtschein, M. Villa, G. Bonera, and F. Borsa, Phys. Rev. B \textbf{13}, 4098 (1976).
\bibitem{Soos} T.~T.~P.~Cheung and Z.~G.~Soos, J. Chem. Phys. \textbf{69}, 3845 (1978).
\bibitem{Notenematic} At lower temperatures, this is not excluded
  though. Indeed it is possible that if a spin-nematic state develops
  at some critical temperature, the 4-spin correlation function does
  slow down, whereas the spin-spin correlation function does not. If this is so, we would predict a strong change of the resonance lineshape.
\bibitem{CepasKagome}  O. C\'epas, C. M. Fong, P. W. Leung, and C. Lhuillier, Phys. Rev. B 78, 140405(R) (2008).
\bibitem{Mendels} For a review, see P. Mendels and F. Bert, J. Phys. Soc. Jpn.  \textbf{79},011001 (2010).
\bibitem{Zorko} A. Zorko, S. Nellutla, J. van Tol, L. C. Brunel, F. Bert,   F. Duc, J. C. Trombe, M. A. de Vries, A. Harrison, and P. Mendels, Phys. Rev. Lett. \textbf{101}, 026405 (2008).
\bibitem{Rousochatzakis} I. Rousochatzakis, S. R. Manmana, A. M. L\"auchli, B. Normand, and F. Mila, Phys. Rev. B \textbf{79}, 214415 (2009).
\end{thebibliography}
\end{document}